\documentclass[aps,prb,reprint,showpacs,superscriptaddress,groupedaddress]{revtex4-1}
\usepackage{graphicx}
\usepackage{amsmath}
\usepackage{amssymb}
\usepackage{dcolumn}
\usepackage{lipsum}
\usepackage{graphicx}
\usepackage{dsfont}
\usepackage{gensymb}
\usepackage{latexsym}
\usepackage{rotating}
\usepackage{bbm}
\usepackage[usenames,dvipsnames]{xcolor}  
\usepackage{float}
\usepackage{epsfig}
\usepackage{psfrag}
\usepackage{natbib}
\usepackage{bm}
\usepackage{eucal}
\usepackage{mathrsfs}
\usepackage{braket}
\usepackage{enumerate}
\usepackage{longtable}
\usepackage{subfigure}
\usepackage{bm}
\usepackage{hyperref}
\usepackage{amsfonts}
\usepackage{dcolumn}
\usepackage{bm}
\usepackage{subfigure}

\newcommand{\be}{\begin{equation}}
\newcommand{\ee}{\end{equation}}
\newcommand{\bn}{\begin{eqnarray}}
\newcommand{\en}{\end{eqnarray}}

\usepackage{color} 

\usepackage{hyperref}
\hypersetup{
colorlinks=true,final=true,
        linkcolor=red,
        citecolor=blue,
        filecolor=blue,
        urlcolor=blue,
}
\begin{document}

\author{Swagata Acharya$^{1}$}\email{swagata@phy.iitkgp.ernet.in}
\author{Dibyendu Dey$^{1}$}
\author{T. Maitra$^{2}$}
\author{A. Taraphder$^{1,3}$}\email{arghya@phy.iitkgp.ernet.in}
\title{The Iso-electronic Series $Ca_{2-x}Sr_{x}RuO_{4}$: Structural Distortion, Effective Dimensionality, Spin Fluctuations and Quantum Criticality}
\affiliation{$^{1}$Department of Physics, Indian Institute of Technology,
Kharagpur, Kharagpur 721302, India}
\affiliation{$^{2}$Department of Physics, Indian Institute of Technology,
Roorkee, Roorkee 247667, India}
\affiliation{$^{3}$Centre for Theoretical Studies, Indian Institute of
Technology Kharagpur, Kharagpur 721302, India}

\begin{abstract}
The iso-electronic $d^{4}$ compounds of the $4d$ series show rich phase diagrams due to competing
spin, charge and orbital degrees of freedom in presence of strong correlations and structural distortions. 
One such iso-electronic series, $Ca_{2-x}Sr_{x}RuO_{4}$, is studied within the GGA (and spin-orbit coupled GGA) plus DMFT formalism using the hybridization expansion of exact low temperature continuous time Quantum Monte Carlo impurity solver. While the local dynamical correlations make $Sr_{2}RuO_{4}$ a Hund's metal, they drive $Ca_{2}RuO_{4}$, the compound at the other end of the series, to a Mott insulating ground state. We study the dynamic and static single-particle and local irreducible vertex-corrected two-particle responses across the series at three different points ($x = 2.0, 0.5, 0.0$) to understand the anomalous cross-over from Hund's metal ($x = 2.0 $) to a Mott insulator ($x = 0 $) and find that a structural distortion is likely to be responsible for the cross-over. Further, dynamical correlations reveal that the band-width ($W$) of the Hund's metal is larger than its effective local Hubbard $U$, and a finite Hund's coupling $J_{H}$ helps it remain in a bad metallic and nearly spin-frozen state over a large temperature range. $Ca_{2}RuO_{4}$, on the other hand, is intrinsically driven to the proximity of a Mott transition due to narrowing of band width ($U/W > 1.5$), though its finite temperature excitations indicate bad metallicity. We show that there is a critical end point of second-order structural transition at $x = 0.5$, where spin fluctuations become critically singular and follow the exact scaling of conformally invariant boundary field theory. We also find that this critical end point of quasi-$3D$ nature is associated with an effective dimensional cross-over between the $x = 2.0$ and $x=0.0$ quasi-$2D$ structures. Finally we draw a modified magnetic phase diagram of the material, showing a fan-like region starting from the quantum critical end point at $x = 0.5.$  
\end{abstract}


\maketitle

Materials with strong electronic correlations~\cite{kotliar,dagotto} can be realized in partially filled d- and f- electron systems. Phenomena like Mott metal-insulator (MIT) transition~\cite{mott}, 
unconventional high-$T_{c}$ superconductivity~\cite{muller,pwa1,pwa2}, colossal magnetoresistance~\cite{gm1,gm2,cm1,tm1} are some of the dramatic ones that arise, solely or partly, due to strong local correlations and can fall in either of the effective single or multi-orbital frameworks depending on the active orbitals at the Fermi level. Correlated materials are realized in many partially filled d- and f- electron systems and the proximity of a Mott transition makes some of these materials rather interesting.
At the same time there are multi-orbital systems such as Ruthenates~\cite{maeno1,ovchini}, 
iron pnictides~\cite{haule,pnic1} and chalcogenides~\cite{sun} which are metals with
strong correlations but are not at the border of a Mott insulating phase. 
The role of Hund's coupling~\cite{hund1,hund2,hund3,hund4,hund5} in single and two-particle responses in many such multi-orbital materials have now been extensively investigated. The Hund's coupling leads to an exponential suppression of the coherence scale of a multi-orbital metal and leads to a large spin-frozen non Fermi-liquid phase. Hund's coupling has profound and distinct effects on spin, orbital and charge degrees of freedom. Nearly degenerate d-orbital systems away from half-filling are driven away from the Mott transition as the Hund's coupling prevents opening of a dynamical charge gap. These disparate, double-faced nature of Hund's coupling in controlling the properties of a correlated metal earned considerable recent interest~\cite{georges-Janus}. 

Ruthenates appear to be tailor-made for investigating the role of Hund's coupling at and away from half-filling. Being 4d-materials, they have less localization effects than their 3d-counterparts and are relatively less strongly correlated (Hubbard $U$ lesser than the bandwidth).This implies that they are relatively far from the Mottness. The symmetry of the Ru $t_{2g}$ orbitals favors a large hybridization with 
$O-p$ orbitals and leading to a large splitting between the $t_{2g}$ and $e_{g}$ orbitals. This, in turn, populates the 3 $t_{2g}$ orbitals with 4 electrons and leads to a substantially lowered spin state than that of isoelectronic Manganites. An extensive of study on the isoelectronic Ruthenates,
$CaRuO_{3}, SrRuO_{3}$~\cite{casr1,casr2,casr3,casr4} and $BaRuO_{3}$~\cite{ba1,ba2}, reveals that all of them can be regarded as prototypical examples of Hund's metal with $U$ values substantially lesser than the respective bandwidths. All these systems, being away from half-filling, are driven far from the proximity 
of Mott transitions by the Hund's coupling $J_{H}$. The role of Van Hove singularity and $J_{H}$ have been extensively studied~\cite{ba1} in these compounds to rationalize their electronic and magnetic properties. Reasonable values of $U$ and $J_{H}$ have been gleaned~\cite{casr2,ba1,ba2} on these materials from experimental single and two-particle features. The ground states and the finite temperature non-Fermi-liquid states are described within three orbital LDA+DMFT framework with such values of $U$ and $J_{H}$. Large mass enhancement factors~\cite{casr2}, substantial increment in the linear specific heat coefficient $\gamma$ and absence of a Mott insulating phase are common to all these three materials. The difference between them, however, concerns the nature of magnetic ground states and $Ru$-$O$-$Ru$ bond angles. The smaller $Ca$ ion leads to a larger rhombohedral distortion ($Ru$-$O$-$Ru$ bond angle is 150\degree) of the lattice than that of the rhombohedral $GdFeO_{3}$ structure of $SrRuO_{3}$ ($Ru$-$O$-$Ru$ bond angle is 163\degree). The bandwidth of $CaRuO_{3}$ reduces substantially and the density of states becomes less at the Fermi level leading to a (non-ferromagnetic) magnetic ground state unlike $SrRuO_{3}$. However, both these materials remain Hund's metals. On the other hand $BaRuO_{3}$ has no $GdFeO_{3}$ distortions with $Ru$-$O$-$Ru$ bond angle 180\degree and is perfectly cubic.

This scenario, however, drastically changes for $d^{4}$ Ruthenates, $Sr_{2}RuO_{4}$ and $Ca_{2}RuO_{4}$. While $Sr_{2}RuO_{4}$ is a $p$-wave superconductor at around 1.5K (though contrary views appeared recently~\cite{Mckenzie,kivelson}) and a Hund's metal at low temperatures, becoming non-Fermi-liquid above $T_{FL}$=25K. The isoelectronic member at the other end of the series, $Ca_{2}RuO_{4}$, is a Mott insulator with an antiferromagnetic (AFM) ground state for $T< 113K$, a paramagnet state for $T < 356K$~\cite{ca2r1} and a bad metal above $356K$~\cite{ca2r2} all the way up to $\sim 1300K$. Recent studies on $Sr_{2}RuO_{4}$ have  established the fact that it is non-Fermi-liquid above 25K, with Curie-Weiss susceptibility, can be explained satisfactorily within Hund's metal framework with $U=2.3eV$ and $J=0.4eV$~\cite{entropy}. While the Hund's metal picture of this compound seems well settled, it fails in case of $Ca_{2}RuO_{4}$~\cite{pavarini}. The $U$ and $J_{H}$ values for this compound needs to be settled within an $LDA+DMFT$ analysis by putting the single and two-particle responses to test against the experimental findings. At the same time it is interesting to investigate why, instead of having all the required features of being a tailor-made Hund's metal (finite and large $J$, not half-filled), the system chooses to become a correlated (Mott) metal. One relevant question is whether the smaller $Ca$ ions lead to a larger distortion of the octahedra. Existing literatures suggest that the strong distortion of $RuO_6$ octahedra, associated with rotation, tilting and flattening respectively, drives the ferro- or anti-ferro-magnetic~\cite{khaliulin} nature of the ground states of $Ca_{2-x}Sr_{x}RuO_{4}$~\cite{friedt} and stabilizes them, while in case of $CaRuO_{3}$ the distortion is comparatively less because there is only one $Ca$ ion in the unit cell, instead of two for $Ca_{2}RuO_{4}$. 
But this ``large distortion" argument immediately raises questions: 
is there an effective dimensional cross-over across the series as one starts replacing $Sr$ by $Ca$? As 
we move across the series ($x=2$ to $x=0$), is some kind of quantum critical point encountered? 
How do the local charge and spin fluctuations evolve across the series? Are they critical at any finite $x$? 
Keeping these questions in mind, we would like to focus on the recently established magnetic phase diagram for the series~\cite{carlo}.
Our aim is to establish a modified magnetic and electronic phase diagram for the series, systematically analyzing the 
structural distortion, dimensional crossover, spin fluctuations and possible aspects of quantum criticality across this series.
Although substantial amount of experiments for probing charge, spin and orbital sectors across the phase diagram have been 
carried out, systematic theoretical studies are lacking. Finally We perform a first-principles calculations (GGA and spin-orbit (SO) coupled GGA) followed by dynamical mean field theory using state of the art ``exact" continuous time quantum Monte-Carlo (CT-QMC) impurity solver. We probe both single- and two-particle vertex-corrected static and dynamic responses for this series by using hybridization expansion of the CT-QMC solver.
We first perform ab-initio density functional theory calculations within GGA and GGA+SO for $Sr_{2}RuO_{4}$ using the full potential linearized augmented plane-wave (FP-LAPW) method as implemented in the WIEN2k code\cite{wien2k}. At the outset we discuss the results from only GGA calculations (without SO). We perform Wannierization of the Wien2k output bands around the Fermi level via interface programs like WANNIER90~\cite{wan}, WIEN2WANNIER~\cite{winwan}. This would, in turn, give us the Wannier orbitals around the Fermi level which serve as inputs of the DMFT self-consistency calculation. 
Similar procedure for the calculations are followed subsequently 
for $Ca_{1.5}Sr_{0.5}RuO_{4}$ and $Ca_{2}RuO_{4}$. However, SO coupled GGA is not be performed for these two compounds 
(the explicit reason is discussed later). From the first-principles calculations, we would also try to address the role of van hove singularities, effective band widths of itinerant bands crossing the Fermi level,
and $Ru$-$4d$ $t_{2g}$-$O$-$2p$ hybridization in tuning the local electronic and magnetic properties of this series. At the next level, using DMFT, we rationalize our choices of $U$ and $J_{H}$ for different $x$, and the role of local correlations in modifying the low energy single- and two-particle dynamic and static responses. 

\section{$Sr_{2}RuO_{4}: GGA+DMFT$}
\begin{figure}
\includegraphics[width=0.98\textwidth]{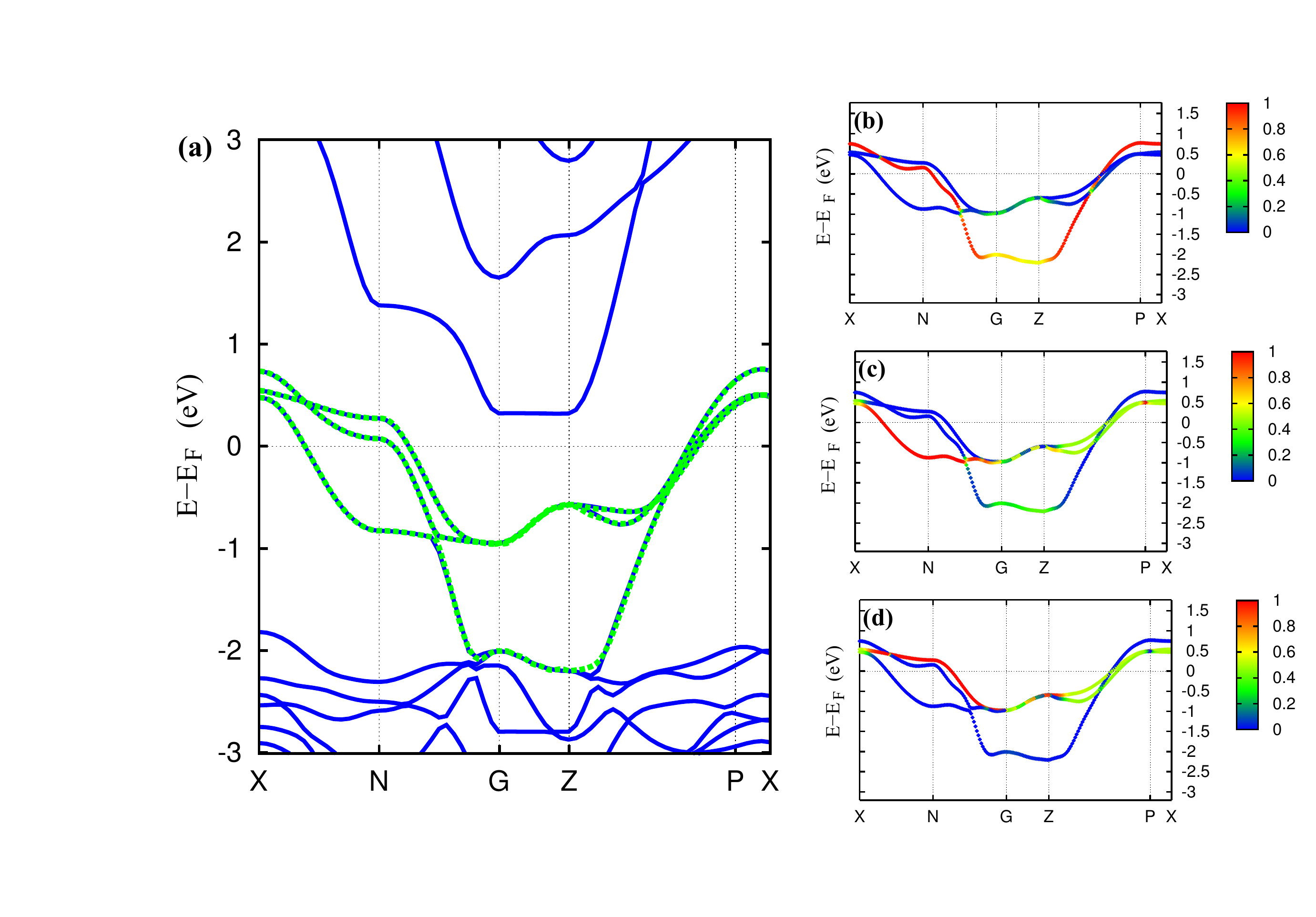}
\caption{(left panel) Band structure for $Sr_{2}RuO_{4}$ and Wannier fitting for the bands crossing the fermi level.
(right panel) Band characterization of the Wannier fit bands for $Sr_{2}RuO_{4}$. The contribution of the $d_{xy}$,
$d_{xz}$ and $d_{yz}$ orbitals respectively to the maximally localized Wannier projected orbitals are shown in (b), (c), (d).}
\label{fig1}
\end{figure}

Experiments suggest that $Sr_{2}RuO_{4}$ is a metal down to 1.5K. The bad metallic nature at high temperatures 
crosses over to a Fermi liquid at $\sim 25K$, followed by a $\sim 1.5K$ superconducting transition. The delocalized $Ru\,4d$-orbitals and a large $Ru$ atomic weight (Z=44) make SO coupling
operative in these materials (as the SO coupling strength $\sim Z^{4}$). Whether a $p$-wave
triplet instability, derived from a momentum-dependent SO coupling, drives a pairing instability in the material at low temperatures, has been debated over last two decades. However, this superconductivity is not the focus of our local DMFT analysis. We are interested in probing the role of SO coupling on single- and two-particle correlated dynamical responses and have gone down to temperatures of order $19K$, well inside the normal phase. We discuss the results we obtained from DFT+DMFT calculations on $Sr_{2}RuO_{4}$. 
The first-principles calculations and subsequent wannierization provide us with 3 orbitals 
active at the Fermi level (Fig~\ref{fig1}). These orbitals are primarily derived 
from $Ru$-$4d$ and have reasonably large hybridization with the $O$-$p$ states at the Fermi level. 
While the hybridization of the $Ru$-$4d$ and apical Oxygens are predominantly 
between $Ru$-$4d$\, $d_{xz}, d_{yz}$ and $O$-$2p$ $p_{x}, p_{y}$\pagebreak 
\onecolumngrid

\begin{figure}
\includegraphics[width=0.98\textwidth]{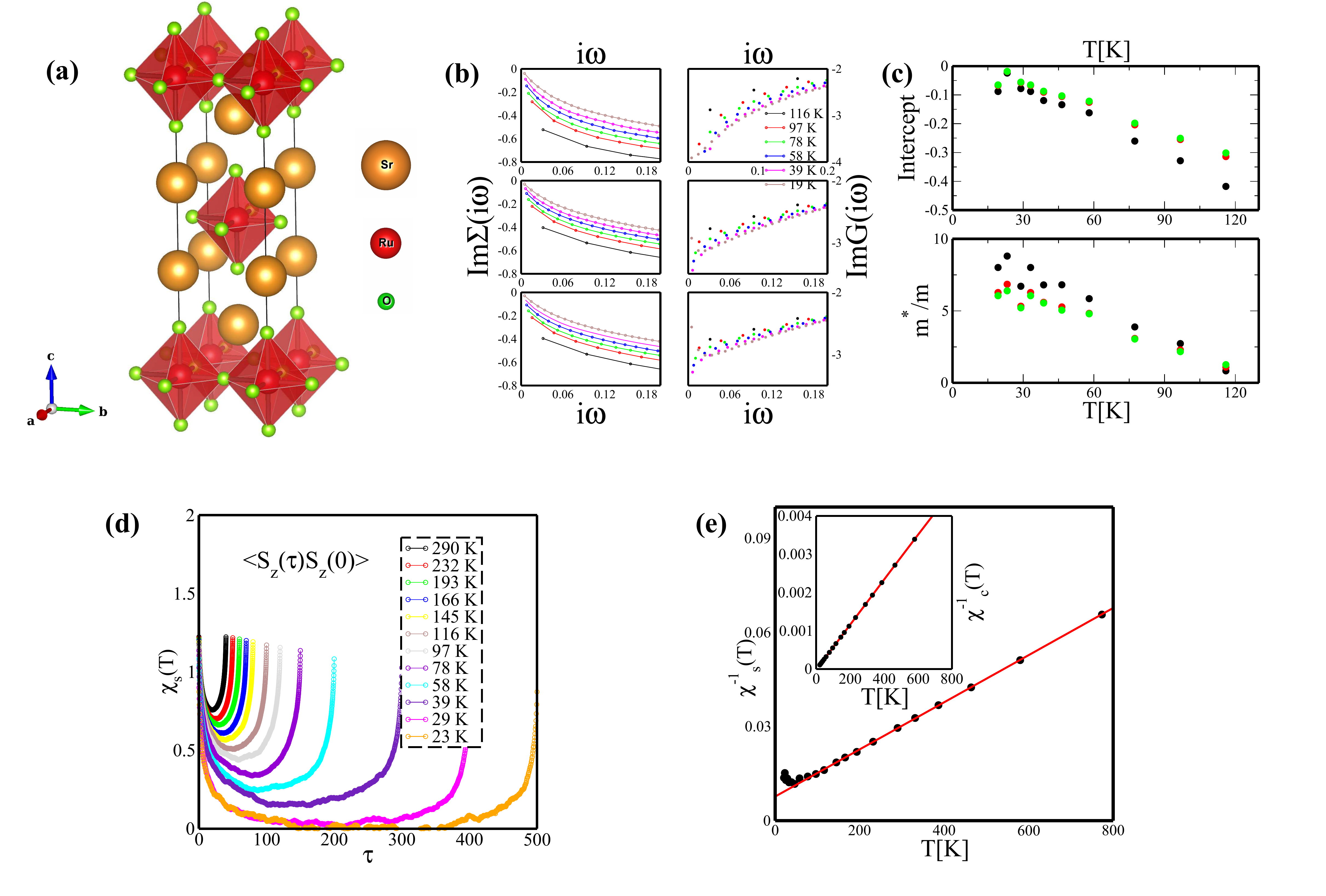}
\caption{(a) Crystal structure of $Sr_{2}RuO_{4}$ with (space-group I4/mmm (139)) high symmetry tetragonal
structure ($a$=3.8606 A$^{0}$ and $c$=12.70658 A$^{0}$).
(b) The Im$\Sigma(i\omega_{n})$ and Im$G(i\omega_{n})$ for
three orbitals ($d_{xy}, d_{xz}, d_{yz}$) over a range of temperatures show retrieval of coherence at lower
temperatures. (c) The intercepts of Im$\Sigma(i\omega_{n})$ at $\omega$=0 and the renormalized mass
enhancement factors $m^{*}/m_{DFT}$ are shown for a large range of temperatures. (d) The dynamic spin
susceptibilities ($\chi_{s,loc}(\tau)$) over a range of temperatures show the tendency towards retrieval of a low temperature
Fermi-liquid phase. (e) The local static spin susceptibility $\chi_{s,loc}(T)$ as a function of temperature shows
the low temperature Fermi-liquid phase sets in at $\sim 41 K$, where the nature of the susceptibility
deviates from singular Curie-Weiss behavior.}
\label{fig2}
\end{figure}

\twocolumngrid

\noindent orbitals, the $p_{y}, p_{z}$ orbitals 
of the Oxygens in the $RuO_{2}$ plane hybridize with the $Ru$\, $d_{xy}$ and $d_{yz}$ orbitals 
respectively. The latter kind of hybridization is more favorable, and is 
reflected in the energy range of 300 meV about the Fermi energy, while the 
earlier one is relevant away from the Fermi level ($\sim$ 1 eV). These hybridizations lead to 
dispersive electrons in the $RuO_{6}$ octahedral geometry. We perform wannierization with these three bands 
(of mixed $O-2p$-$Ru-4d$ character) and use the Wannier fit to perform the DMFT. In Fig.~\ref{fig1} the band 
characterization of the Wannier bands are shown. The three figures 
Fig.~\ref{fig1}(b), (c) and (d) respectively characterize the $d_{xy}$, $d_{xz}$ and $d_{yz}$ contributions to the bands crossing the Fermi level. Although the rotation and mixing of the original bands have taken place, the band characterization, as is apparent in the Fig~\ref{fig1}, allows us to refer to them as nominally $d_{xy}$, $d_{xz}$, $d_{yz}$ bands  respectively. Subsequently we will take these Wannier bands as input of DMFT and refer to them similarly for the rest of the discussion. We choose a value $U$=2.30 eV and $J_{H}$=0.40 eV, as prescribed in literature~\cite{entropy} for $Sr_{2}RuO_{4}$. The choice is motivated by the fact that it predicts the experimentally derived values of local moment and quasi-particle weight over a large range of temperatures. Imaginary parts of the single-particle self-energy Im$\Sigma(i\omega_{n})$ (Fig~\ref{fig2}(b)) and the Green's function Im$G(i\omega_{n})$ are plotted for individual orbitals over a range of temperature between 120K to 19K. Two things are clearly visible from the results: the intercepts of Im$\Sigma(i\omega_{n})$ at $\omega$=0 monotonically reduces (Fig~\ref{fig2}(c)) and approaches zero and Im$G(i\omega_{n})$ becomes more coherent as the temperature is lowered. The systematic retrieval of coherence is already evident from Im$\Sigma(i\omega_{n})$, and Im$G(i\omega_{n})$. We fit Im$\Sigma(i\omega_{n})$ to a fourth order polynomial in $i\omega_{n}$. We find that the mass enhancement, which is related to the coefficient ($c_{1}$) of the linear term in the expansion $m^{*}/m = 1+c_{1}$, increases with lowering of temperature (Fig~\ref{fig2}(c)). It can be seen that the $xy$ orbital has a larger mass enhancement than the $xz$ and $yz$ orbitals, and concomitantly the intercepts of Im$\Sigma(i\omega_{n})$ at $\omega$=0 are larger for $xy$ orbital than the other two (Fig~\ref{fig2}(c)). The $m^{*}/m$ for $d_{xy}$ is 5.8 and for $d_{xz},\, d_{yz}$ are 4.5 in excellent agreement with experiments and \pagebreak

\onecolumngrid

\twocolumngrid
\noindent earlier theoretical studies. On the other hand, from these results, the non-Fermi-liquid to Fermi-liquid cross-over that is observed in experiments around 25K, cannot be demonstrated cleanly. We switch over to static and dynamic vertex corrected spin and charge susceptibilities to glean whether the incoherence to coherence
cross-over is reflected in the two-particle sector: it often happens that a low energy single-particle description might not be adequate to trace such cross-over scales associated with multi-particle dynamics. The local dynamic spin susceptibilities ($\chi_{s,loc}(\tau)$) have been plotted against $\tau$ (Matsubara time) for a large range of temperatures (Fig~\ref{fig2}(d)). We also find out the time integrated static local spin susceptibility ($\chi_{s,loc}(T)$) (Fig~\ref{fig2}(e)). $\chi_{s,loc}(\tau)$ below $\sim 41K$ seems to have a zero intercept at $\tau=\beta/2$ and a $\tau^{2}$ behavior around $\tau=\beta/2$, while for higher temperatures the intercept is finite and large and increases with a rise in temperature. $\chi_{s,loc}(\tau)$ also deviates from a low-energy $\tau^{2}$ behavior with increasing temperatures. To add to that, $\chi_{s,loc}(T)$ clearly shows a deviation from its high temperature Curie-Weiss behaviour at $\sim 41K$. Although the low temperature behavior of $\chi_{s,loc}(T)$ below 41K is not Pauli-like, indicating that the spins are not quenched completely below this temperature, the strong deviation from Curie-Weiss behavior is a signature of emergence of a low temperature coherence scale. The dynamic local charge and orbital susceptibilities ($\chi_{c,loc}(\tau)$, $\chi_{o,loc}(\tau)$), however, remain more singular than the spin susceptibility down to the 19K, and while the spin fluctuations are quenched partially, the static local charge and orbital susceptibilities ($\chi_{s,loc}(T)$, $\chi_{o,loc}(T)$) remained Curie-Weiss-like down to 19K. The consequences of the same will be discussed in a separate study and we skip it for now. However, experimentally, the orbital degrees of freedom also seem to get quenched partially at lower temperatures which is not reproduced in the present analysis. An account of this may need the inclusion of SO coupling within the first-principles calculations. 
\pagebreak
\onecolumngrid

\begin{figure}
\includegraphics[width=0.98\textwidth]{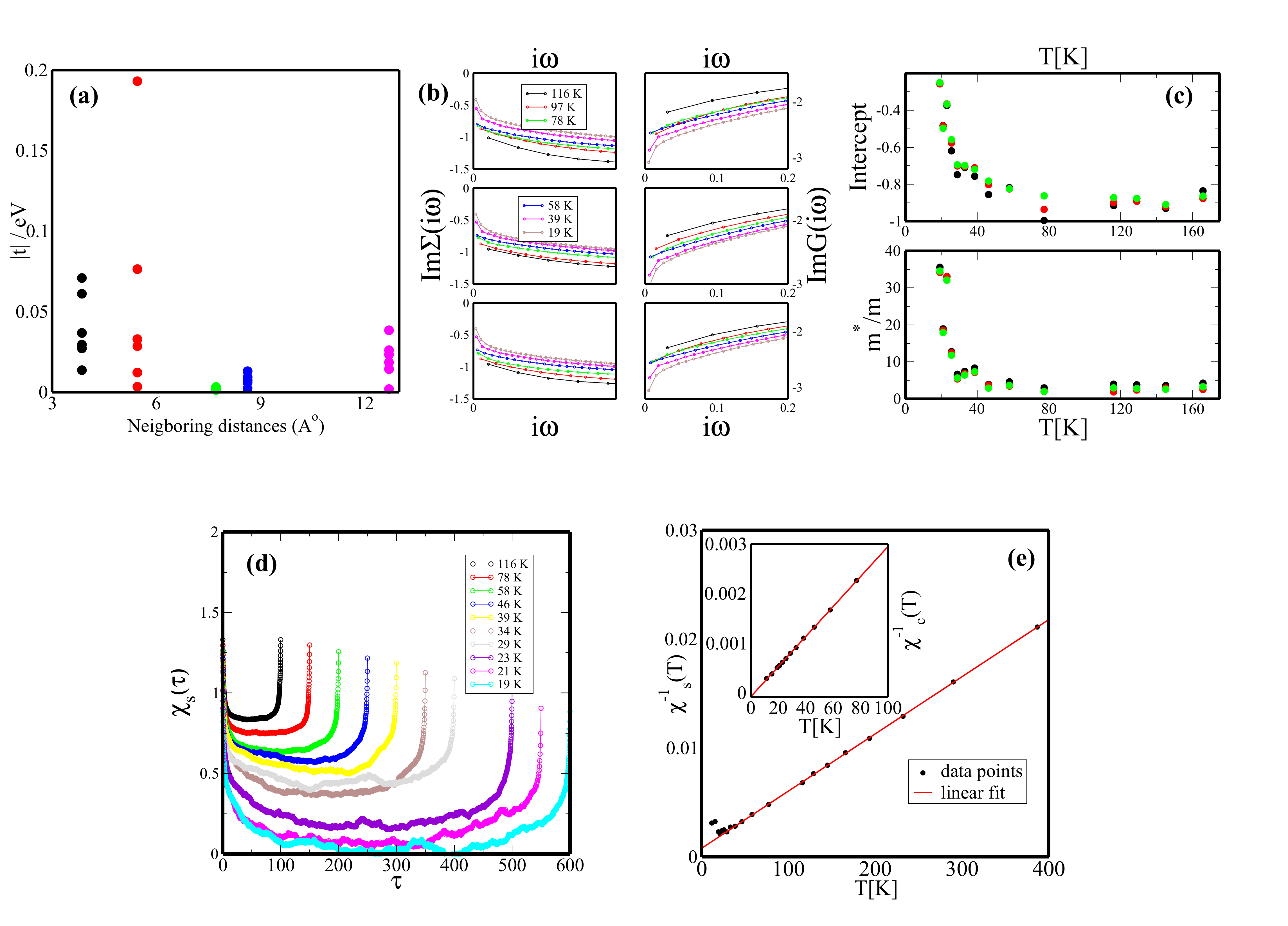}
\caption{(a) The in-plane ($t_{ab}$) and out of plane ($t_{c}$) hopping elements
for $Sr_{2}RuO_{4}$, extracted from real space Wannier hopping matrix showing
the $2D$ electron itinerant nature ($a$=3.8606 A$^{0}$ and $c$=12.70658 A$^{0}$) (b) The Im$\Sigma(i\omega_{n})$ and Im$G(i\omega_{n})$ for three orbitals ($d_{xy}, d_{xz}, d_{yz}$) are shown
over a range of temperatures. (c) The intercepts of Im$\Sigma(i\omega_{n})$ at $\omega$=0 and the
renormalized mass enhancement factors $m^{*}/m_{DFT}$ show orbital specific retrieval of coherence at low temperatures.
(d) The dynamic spin susceptibilities ($\chi_{s,loc}(\tau)$) for a range of temperatures show tendency towards
retrieval of a low temperature Fermi-liquid phase.
(e) The local static spin susceptibility $\chi_{s,loc}(T)$
shows the low temperature Fermi-liquid phase sets in at $\sim 23 K$,
where the nature of the susceptibility deviates from singular Curie-Weiss behavior.}
\label{fig3}
\end{figure}
\twocolumngrid
\noindent A significant lifting of orbital degeneracy can suppress the orbital moment at low temperatures. From a GGA+SO calculations we find that the orbital degeneracies are lifted at different $k$-points across the Brillouin zone by different amounts. The maximum value of SO splitting is $\sim 130$ meV while it is negligible at some symmetry points. The effect of SO, however, can hardly be realized by looking at the momentum integrated density of states of $Sr_{2}RuO_{4}$ across the Fermi level. We find that the nature of $O-p$ and $Ru-d$ hybridization remains more or less similar to the results without SO. We identify the bands active at the Fermi level and perform the wannierization. Similar to our previous calculations, three Wannier fit orbitals have predominant contributions from $Ru-d_{xy}, d_{xz}, d_{yz}$ orbitals. We perform our DMFT calculations
using these SO-coupled orbitals. 

\begin{figure}[h!]
\centering
\includegraphics[width=0.48\textwidth]{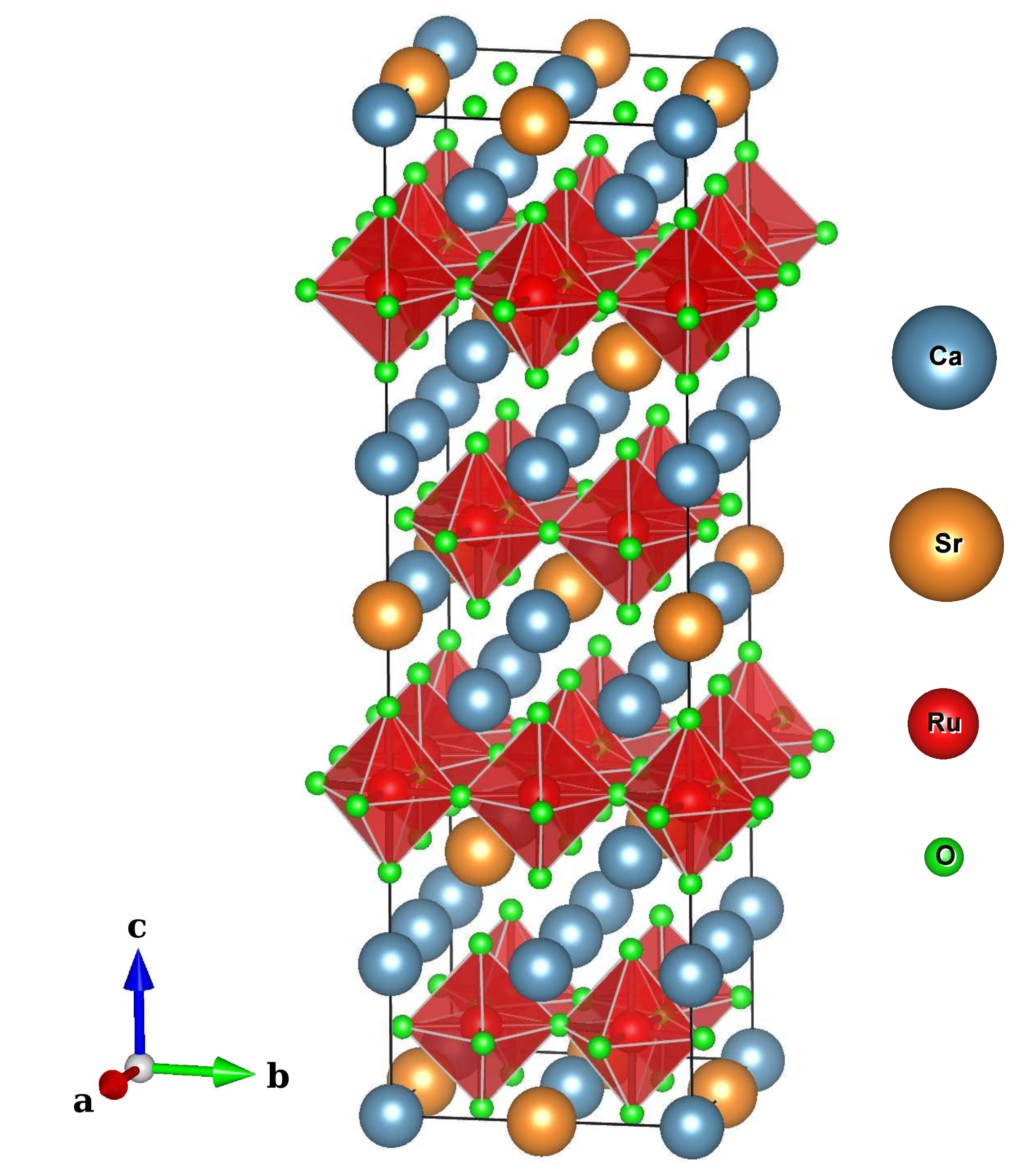}
\caption{The crystal structure of $Ca_{2-x}Sr_{x}RuO_{4}$ with high symmetry (space-group I4/mmm $139$) tetragonal structure ($a$=7.52 A$^{0}$ and $c$=24.1645 A$^{0}$).}
\label{fig4}
\end{figure}

With SO included, the DMFT self-energies (Fig~\ref{fig3}(b)) have similar qualitative features as without SO, though the details are significantly different. Here again the intercept of Im$\Sigma(i\omega_{n})$ (Fig~\ref{fig3}(c)) at $\omega$=0
decreases with increasing temperature, but the drop is sharper and non-monotonic across 24K. The $m^{*}/m$ is, in turn, found to increase sharply across it (Fig~\ref{fig3}(c)). Both Im$\Sigma(i\omega_{n})$ and Im$G(i\omega_{n})$  can be found to have sharp change of slope below $i\omega_{n}$=0.03 eV (Fig~\ref{fig3}(b)), marking the advent of coherence below T$ = 24K$. At this point we look at the two-particle vertex corrected responses. The $\chi_{s,loc}(\tau)$ (Fig~\ref{fig3}(d)) shows Fermi-liquid features, with a $\tau=\beta/2$ intercept $\sim 0$ and $\tau^{2}$ behavior at low energies, below 24K. The $\chi_{s,loc}(T)$ (Fig~\ref{fig3}(e)) seems to be deviating from the high temperature Curie-Weiss behavior precisely below the same scale of $T$, marking the emergence of coherence in the system. Hence, the scenario with SO is distinctly different from the scenario without SO. With SO included, both the single and two-particle static and dynamic responses register the emergence of coherence quite consistently. The local orbital dynamic susceptibilities get suppressed at low temperatures in the present analysis and become significantly smaller than the local spin susceptibilities. The lifting of degeneracy in the orbital sector due to SO coupling is responsible for the quenching of orbital moment. The dynamic local charge susceptibilities, however, still remain divergent and singular down to 19K implying a soft charge fluctuations down to the lowest temperatures. The values of real space hopping we get from wannierization are also suggestive. The relative magnitudes of the in-plane and out-of-plane hopping elements provide insight into the effective dimensionality of the material. The $c/a$ ratio for $Sr_{2}RuO_{4}$ is 3.294 and the itinerant electrons have the in-plane hopping elements nearly an order of magnitude larger than their out of plane counterpart (Fig~\ref{fig3}(a)). The system is quasi-$2D$ in this case, supported also by a large resistivity anisotropy ratio~\cite{maenoRMP} ($\rho_{c}/\rho_{ab}\sim 200 $) and lesser hybridization between the apical Oxygens and $Ru\,4d\, t_{2g}$ orbitals. It would be interesting to see whether the replacement of $Sr$ by a smaller cation $Ca$ leads to any kind of dimensional cross-over, and whether that cross-over encounters a structural or magnetic critical point.
\section{$Ca_{1.5}Sr_{0.5}RuO_{4}: GGA + SO + DMFT$}
\begin{figure}
\includegraphics[width=0.98\textwidth]{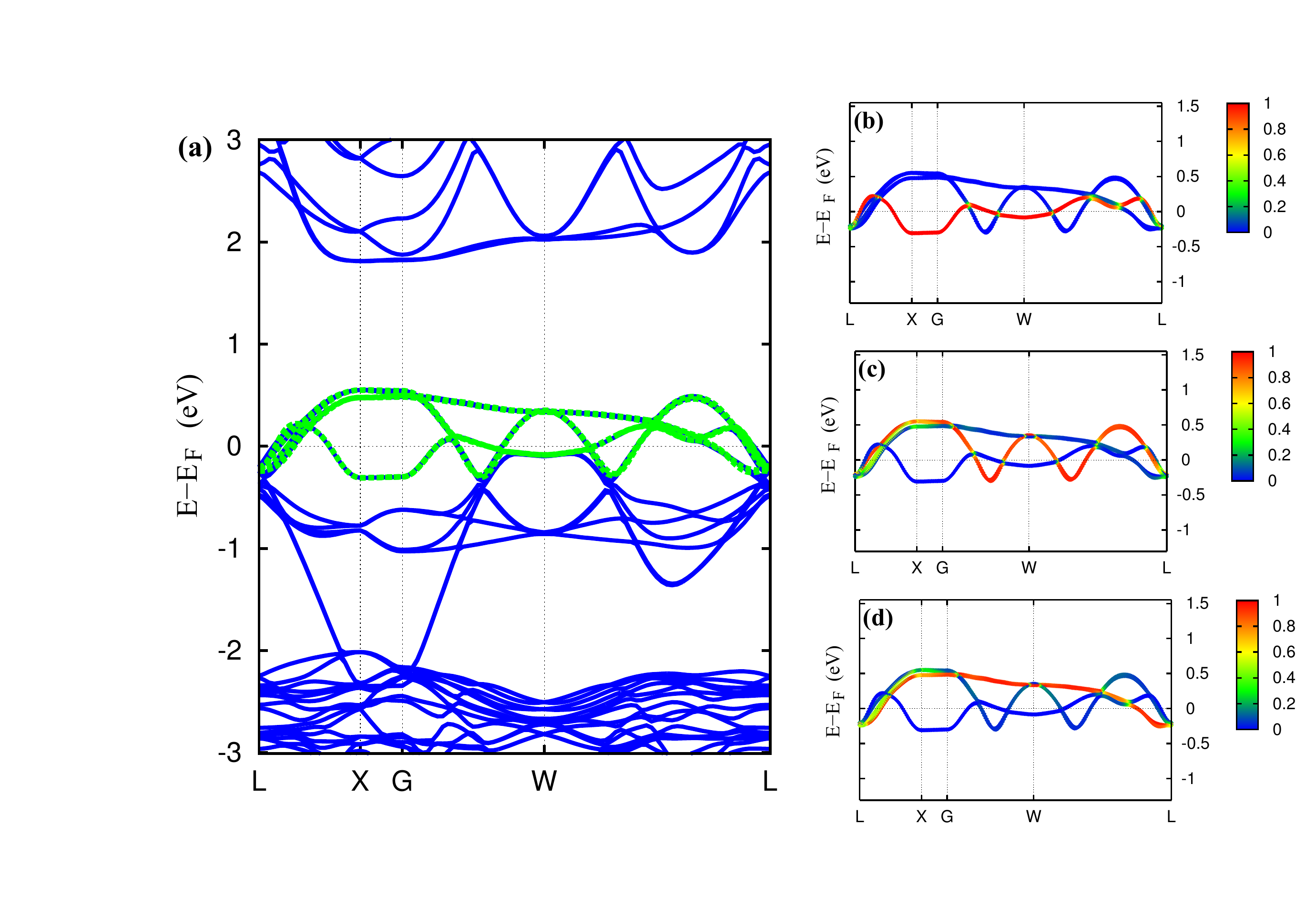}
\caption{(left panel) Band structure for $Ca_{1.5}Sr_{0.5}RuO_{4}$ and Wannier fitting for the bands crossing the fermi level.
(right panel) Band characterization of the Wannier fit bands for $Ca_{1.5}Sr_{0.5}RuO_{4}$. The contribution of the $d_{xy}$,
$d_{xz}$ and $d_{yz}$ orbitals respectively to the maximally localized Wannier projected orbitals are shown in (b), (c), (d).}
\label{fig5}
\end{figure}

$RuO_{6}$ octahedron gets distorted (Fig~\ref{fig4}) significantly as the smaller $Ca$ cation replaces $Sr$. As suggested from experiments~\cite{terakura1,terakura2}, the distortion takes place in steps: first a rotation of the octahedra about the $c$ axis takes place and then a tilt of the octahedra, followed finally by a flattening of it. $Ca_{1.5}Sr_{0.5}RuO_{4}$ undergoes the first two kinds of distortions, starting from a nearly clean undistorted $RuO_{6}$ octahedra in $Sr_{2}RuO_{4}$. 
The GGA calculations allow us to see the primary changes in its band structure thereon. The $Ru$-$4d$
bands get significantly narrowed in this case and the $e_{g}$ orbitals contribute finite density of states at the Fermi level alongside the $t_{2g}$ orbitals. Although the contribution from $e_{g}$ is much less than $t_{2g}$, it is finite and not negligible compared to $t_{2g}$-like density of states at the Fermi level for $Sr_{2}RuO_{4}$. The $d_{xy}$ orbital is the majority contributor to DOS, with a sharp Van-Hove like feature within an energy range of 50 meV of the Fermi level. The $d_{yz}$ and $d_{zx}$ orbitals are secondary contributors. While $p_{y}$ and $p_{z}$ orbitals of in plane ($RuO_2$ plane) Oxygens hybridize strongly with the $Ru\, 4d$ at the Fermi level, $p_{x}$ and $p_{y}$ orbitals of apical Oxygen hybridizes strongly with the $Ru\, 4d$ nearly 700-800 meV away from the Fermi level. However, Wannierization and band characterization suggest that the Wannier fit bands can be identified (similar to $Sr_{2}RuO_{4}$) as predominantly $d_{xy}$, $d_{xz}$, $d_{yz}$-derived as shown in the band characterization plot in Fig~\ref{fig5}. The sharp Van-Hove feature close to the Fermi surface and significantly narrow band-width of the $Ru \, d$ orbitals already place the material close to a ferromagnetic Stoner instability.

\begin{figure}
\includegraphics[width=0.98\textwidth]{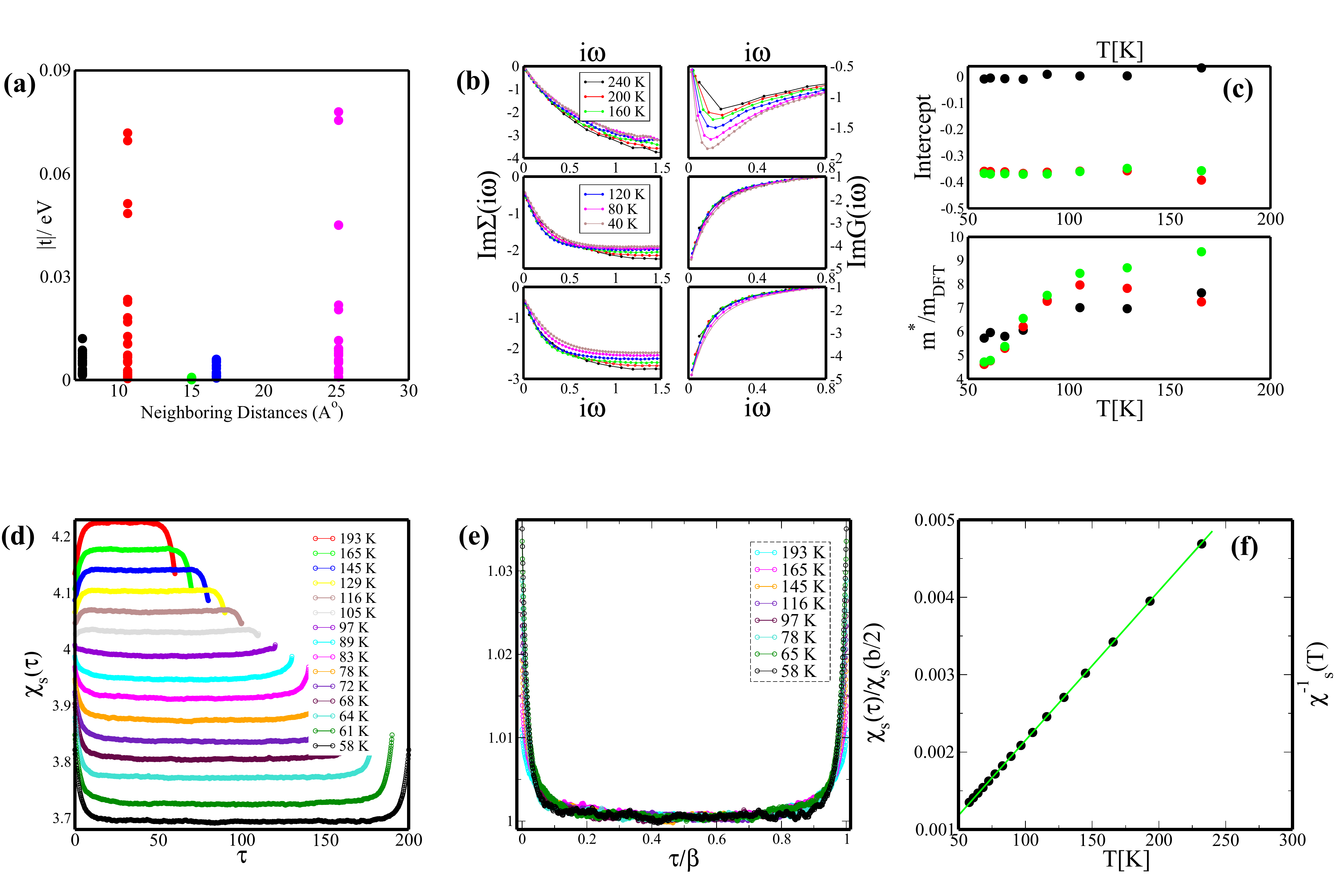}
\caption{(a) The in-plane ($t_{ab}$) and out of plane ($t_{c}$) hopping elements for $Ca_{1.5}Sr_{0.5}RuO_{4}$, extracted from real space Wannier hopping matrix showing the $3D$ electron itinerant nature. (b) The Im$\Sigma(i\omega_{n})$ and Im$G(i\omega_{n})$ for three orbitals ($d_{xy}, d_{xz}, d_{yz}$) are shown
over a range of temperatures. (c) The intercepts of Im$\Sigma(i\omega_{n})$ at $\omega$=0 and the
renormalized mass enhancement factors $m^{*}/m_{DFT}$ shows orbital selective coherence for a large range of temperatures. (d)
The dynamic spin susceptibilities ($\chi_{s,loc}(\tau)$) for a range of temperatures show that the system is a bad metal
over the complete temperature range as the $\tau/\beta$=0.5 intercepts of different $\chi_{s,loc}(\tau)$ curves are finite
and large. (e) The strong non-Fermi liquid character of the system is evident as the $\chi_{s,loc}(\tau/\beta)$ shows
clear thermal scaling collapse. (f) The local static
spin susceptibility $\chi_{s,loc}(T)$ shows that the system is
non-Fermi-liquid down to the lowest accessible temperatures.}
\label{fig6}
\end{figure}

We use these three Wannier-fit (projected using the $dt_{2g}$ bands) bands as inputs for DMFT. The DMFT  calculations are performed with $U=3.0 eV$ and $J_{H}=0.6 eV$. The rationale behind these parameter values is again the same $-$ the local static quantities are reproduced for a range of
temperatures with these parameters. These numbers are larger than the corresponding numbers for $Sr_{2}RuO_{4}$, for the bandwidth of the bands crossing the Fermi level has decreased significantly now due to smaller cationic distortion. Im$\Sigma(i\omega_{n})$ and Im$G(i\omega_{n})$ are plotted as functions of $i\omega_{n}$ in Fig~\ref{fig6}(b) and they show tendencies towards enhanced coherence with lowering temperatures. However, unlike $Sr_{2}RuO_{4}$, here the material remains non-Fermi-liquid down to the lowest temperatures probed. The intercept,
as shown in Fig~\ref{fig6}(c) of Im$\Sigma(i\omega_{n})$ at $i\omega_{n}$=0 is nearly independent of temperature and remains large and finite down to the lowest temperature. The orbital selective nature of the Im$\Sigma(i\omega_{n})$ can be gleaned now from the fact that $d_{xy}$ orbital has a very small intercept at $i\omega_{n}$=0 compared to $d_{xz}$ and $d_{yz}$ orbitals, but the mass enhancement factors for the $d_{xz}$ and $d_{yz}$ orbitals steadily decrease with lowering temperature (Fig~\ref{fig6}(c)). The dependence is not as monotonic for $d_{xz}$ orbital. This can again be corroborated by the orbital-specific Im$G(i\omega_{n})$:  $d_{xz}$ and $d_{yz}$ orbitals have metallic coherent features while the $d_{xy}$\pagebreak 
\onecolumngrid

\twocolumngrid

\noindent orbital is semi-metallic with a pseudo-gap energy scale that starts getting lower with lowering temperatures. 
We probe the two-particle dynamics and static spin and charge responses for $Ca_{1.5}Sr_{0.5}RuO_{4}$ below.

We note an interesting fact that $\chi_{s,loc}(\tau)$ (Fig~\ref{fig6}(d)) has a completely different functional dependence on $\tau$ at high temperatures (between 600K to 120K) and at lower temperatures (below 120K).
The high temperature $\chi_{s,loc}(\tau)$ suggests that even at low energies the spin is completely
unquenched, while partial quenching of spins at low energies begins below 100K. Thereafter, as the temperature is lowered, the spin quenching increases, but $\chi_{s,loc}(T)$ (Fig~\ref{fig6}(f)) suggests that the
response remains Curie-Weiss-like down to the lowest temperatures that we could access 
within our CT-QMC analysis. Two important observations may be noted at this point. 
Firstly, the $\chi_{s,loc}(\tau)/\chi_{s,loc}(\beta/2)$ follows a beautiful 
scaling (Fig~\ref{fig6}{e}) as a function of $\tau/\beta$ for the temperature range 
between 120K to 30K, with the functional form $Sin(\pi\tau/\beta)^-{(1-\alpha)}$ and 
the scaling exponent $\alpha$=0.10. And secondly, $\chi_{s,loc}(T)$ (Fig~\ref{fig6}(f)) for this material, 
in this temperature range, is more than an order of magnitude larger~\cite{friedt} 
than $\chi_{s,loc}(T)$ for $Sr_{2}RuO_{4}$. And finally, the linear fit to 
the $\chi^{-1}_{s,loc}(T)$ vs $T$ curve shows a $\theta_{c}$ 
(from $\chi^{-1}_{s,loc}(T)=T+\theta_{c}$) which is positive and large (41 K). 
All these signatures unambiguously indicate a strong quantum critical ferromagnetic fluctuation arising from a temperature which is possibly lower than our reach within CT-QMC at $x=0.5$. This could well be a critical end point ($T = 0, x= 0.5$) as there is no experimentally reported structural transition either suppressing or facilitating the critical ferromagnetic fluctuation at this point~\cite{critical}. The orbital selective pseudogapped~\cite{si2} nature of the density of states and non-Fermi-liquid self-energy response also act as supportive informations for the observation of quantum criticality, which has been rigorously shown to be existing in different contexts~\cite{mark,si1,jpcs,INS}in literature, with similar single-particle responses.  

Both the field variables $ImG(\tau)/ImG(\beta/2)$ and $\chi_{s,loc}(\tau)/\chi_{s,loc}(\beta/2)$ 
scale as functions of $\tau/\beta$ which suggests that temperature is the only scale in this parameter
regime. The scaling functional form $Sin(\pi\tau/\beta)^-{(1-\alpha)}$, 
\pagebreak

\onecolumngrid

\twocolumngrid

\noindent hints that a conformally invariant
boundary field theory is a good field theory for $Ca_{1.5}Sr_{0.5}RuO_{4}$, and hence the criticality is strictly local.
At this point we again choose to probe the relative scales of in-plane and out-of-plane hoppings
and find that they have not changed significantly at all from the parent compound $Sr_{2}RuO_{4}$.
The $c/a$ ratio turns out to be $3.30$ which is nearly similar at $x=2.0$.
What interests us here, is the ratio of the in-plane and out-of-plane hopping scales. 
The out-of-plane hopping along the $z$ direction is not suppressed at all; both in-plane and out-of-plane components (Fig~\ref{fig6}(a)) are nearly
of the same order of magnitude. The effective three dimensional nature of the electron hopping suggests that 
in spite of having a similar $c/a$ ratio the $x=0.5$ material is $3D$ in electron itineracy, unlike its $x=2.0$ counterpart.
It's worth noting, at this point, that $c/a$ ratio can not be the lone parameter 
deciding the effective dimensionality of a bulk single crystal. The distortion-induced changes in the $Ru \, d$ and apical $Op$
hybridization scales are equally important in determining the effective dimensionality.
This effective dimensionality, critical ferromagnetic spin fluctuations, finite temperature Curie-Weiss spin susceptibility, 
enhanced positive $\theta_{c}$, finite temperature scaling collapse of dynamic spin susceptibilities, and absence of
any magnetic, orbital or charge ordering down to lowest temperatures suggest that $x=0.5$ is critical.     
This is again of paramount interests, as it is already reported in literature~\cite{critical}
that there is a $T-x$ line that separates the $Ca_{2-x}Sr_{x}RuO_{4}$ phase diagram in to two regions; one
with two fold in-plane susceptibility-anisotropy (at lower $x$ below $x$=0.5)~\cite{nakatsuji} and the other region
for $x > 0.5$ without any anisotropy. This suggests that there is already reasonably compelling evidence that
this $x = 0.5, T = 0$ could well be a quantum critical end point of the second-order structural transition.
Having observed the criticality at $x=0.5$, it would be worth seeing, what is the effect of the 
critical fluctuations at lower $x$ values. We choose to probe $Ca_{2}RuO_{4}$ on a similar line.  
\section{$Ca_{2}RuO_{4}-LPBCA : GGA + DMFT$}

\begin{figure}
\centering
\includegraphics[width=0.40\textwidth]{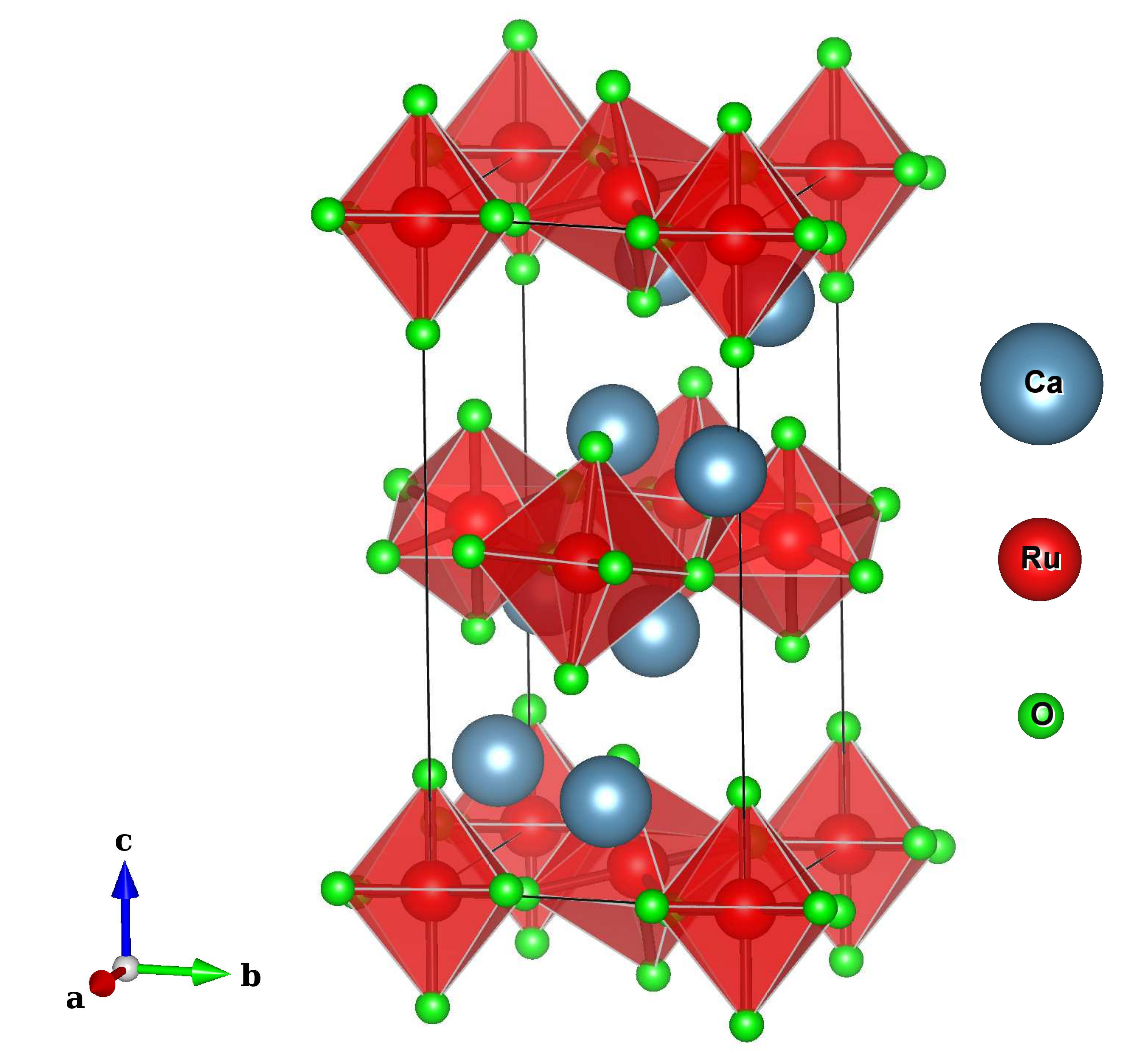}
\caption{Crystal structure of $Ca_{2}RuO_{4}$ with lowered symmetry L-Pbca structure $a$=5.3869 A$^{0}$, $b$=5.6334 A$^{0}$,
$c$=11.7349 A$^{0}$.}
\label{fig7}
\end{figure}

\begin{figure}
\includegraphics[width=0.98\textwidth]{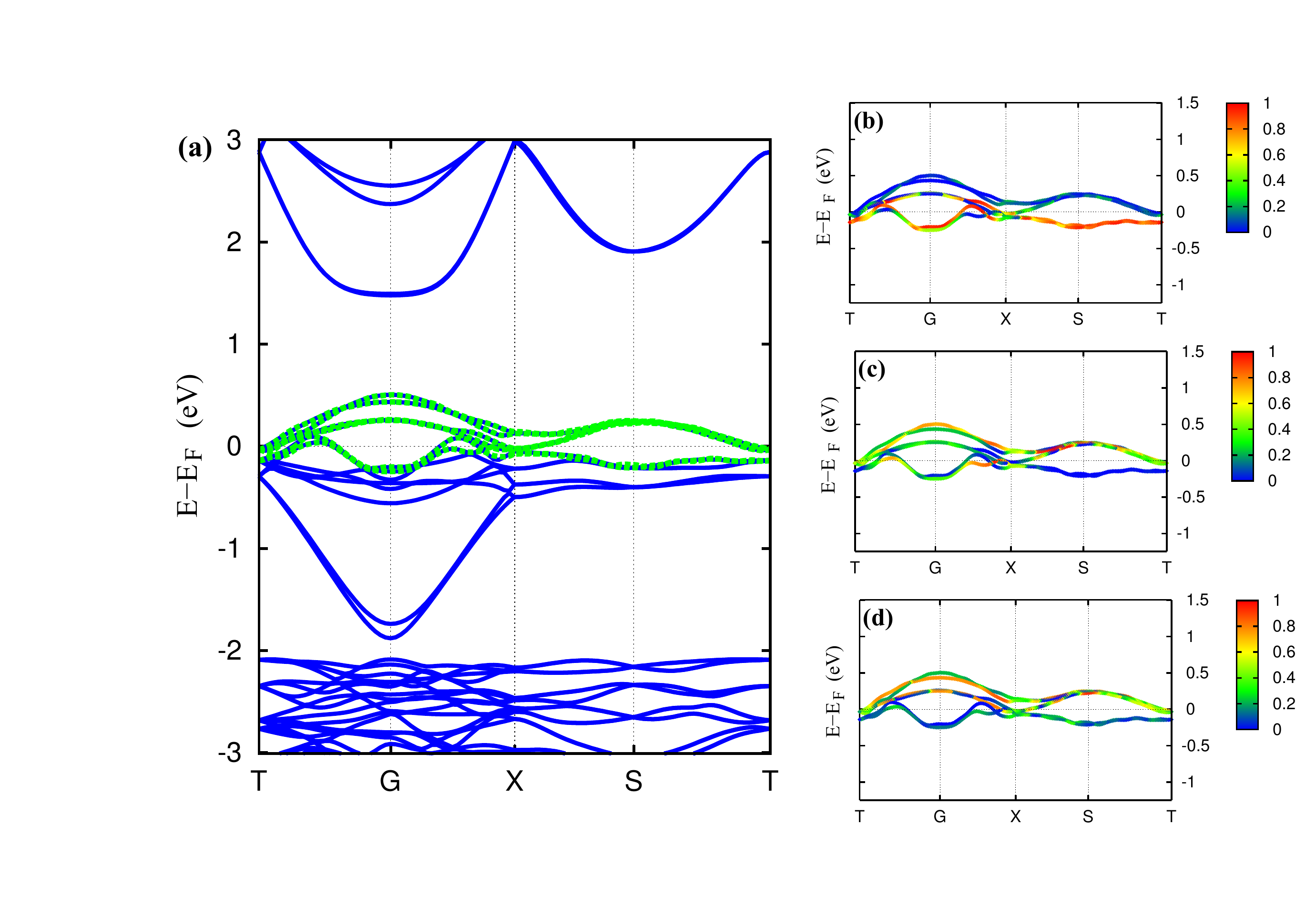}
\caption{(left panel) Band structure for $Ca_{2}RuO_{4}$ and Wannier fitting for the bands crossing the fermi level.
(right panel) Band characterization of the Wannier fit bands for $Ca_{2}RuO_{4}$. The contribution of the $d_{xy}$,
$d_{xz}$ and $d_{yz}$ orbitals respectively to the maximally localized Wannier projected orbitals are shown in (b), (c), (d).}
\label{fig8}
\end{figure}

\begin{figure}
\includegraphics[width=0.98\textwidth]{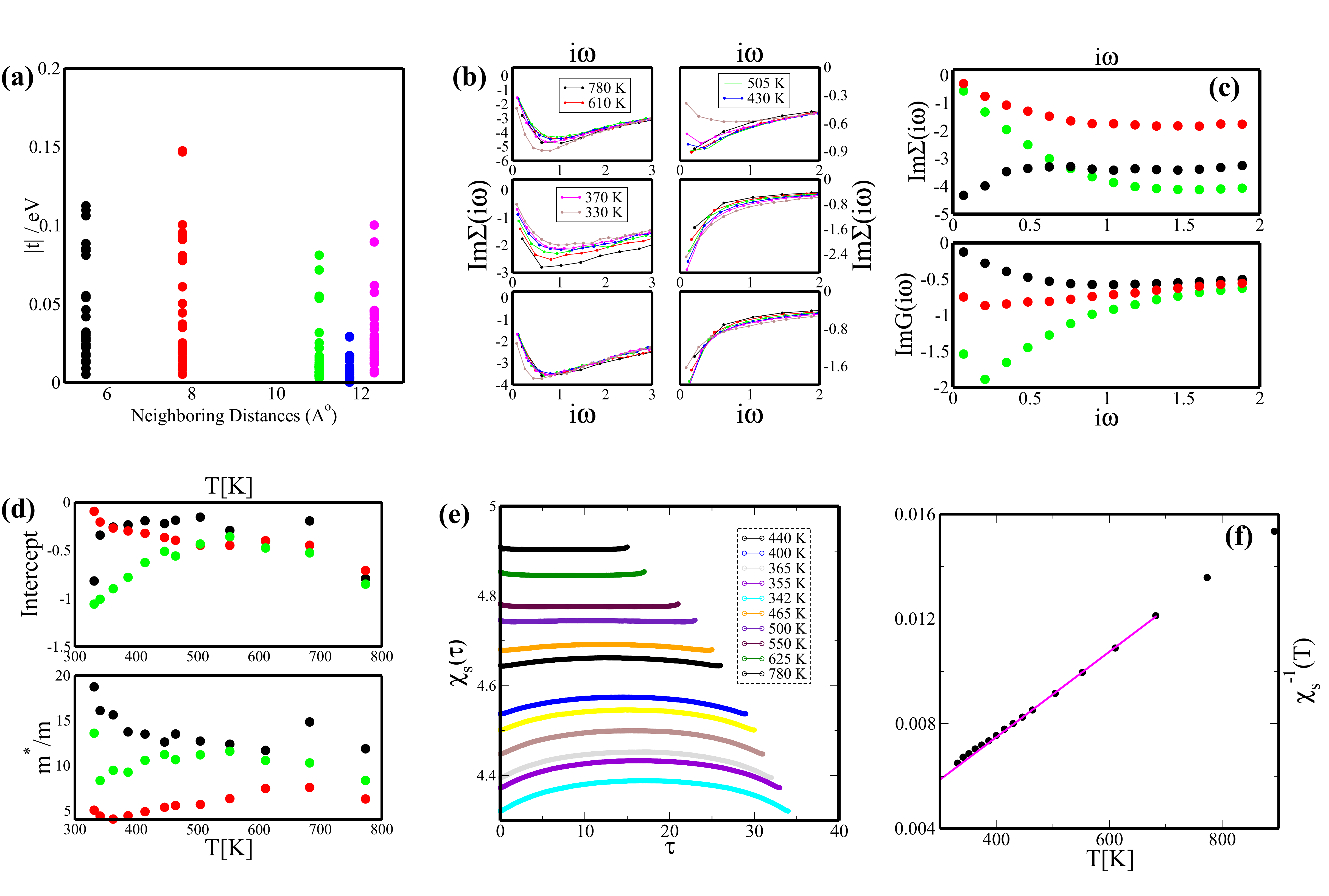}
\caption{(a) The in-plane ($t_{ab}$) and out of plane ($t_{c}$) hopping elements for $Ca_{2}RuO_{4}$,
extracted from real space Wannier hopping matrix showing $2D$ nature of the itinerant electrons.
(b) The Im$\Sigma(i\omega_{n})$ and Im$G(i\omega_{n})$ for three orbitals ($d_{xy}, d_{xz}, d_{yz}$) are shown
over a range of temperatures showing orbital selective Mott transition. (c) The intercepts of Im$\Sigma(i\omega_{n})$ at $\omega$=0 and
the renormalized mass enhancement factors $m^{*}/m_{DFT}$ are shown for a large range of temperatures. (d) The dynamic spin
susceptibilities ($\chi_{s,loc}(\tau)$) show that the system is a
non-Fermi-liquid metal over the complete temperature range shown.
(e) The local static spin susceptibility $\chi_{s,loc}(T)$ shows that the system is
non-Fermi-liquid down to the Mott critical temperature (356 K). However, the behavior deviates from Curie-Weiss above 356 K
and is perfectly Curie-Weiss at higher temperatures.}
\label{fig9}
\end{figure}

As experimental evidences suggest~\cite{nakatsuji,friedt}, $Ca_{2}RuO_{4}$ has an AFM Mott 
insulating state below 113K. Between 356K and 113K it is a paramagnetic Mott insulator and above 356K, 
a paramagnetic bad metal. The metal to Mott insulator transition at 356K is associated with a structural 
transition from a L-Pbca (Fig~\ref{fig7}) to S-Pbca structure. We first do GGA calculations using L-Pbca structural inputs from experiments, perform Wannierization, apply local correlations within DMFT scheme on Wannier-fit orbitals (Fig~\ref{fig8}) and look for a Mott metal-insulator transition. 
While all the $d$ orbitals have substantial contributions to the DOS at the Fermi level, $d_{yz}, d_{xy}, d_{z^{2}}$ have the dominant contributions. There are sharp Van-Hove features near the Fermi level, but
they are farther from the Fermi level than those in $Ca_{1.5}Sr_{0.5}RuO_{4}$. However, the bandwidth for the dispersive $d$ orbitals are even lower in this case than the $x=0.5$ material. The highly distorted octahedra allows for strong hybridization between all $Ru \, d$ orbitals and $O \, p$ orbitals: the apical Oxygen $p_{x}, p_{z}$ and in-plane Oxygen $p_{x}, p_{y}, p_{z}$ have sufficient hybridization with $Ru \, d$ orbitals. All these facts put together, the L-Pbca intrinsically has the tendency towards nesting and thence antiferromagnetic fluctuations that can stabilize a low temperature AFM ground state with a
gain in energy $\sim t^{2}/U$.  
But the L-Pbca structural details do not explain why it should become a Mott
insulator below 356K. It is often suggested in the literature~\cite{pavarini} that the L-Pbca 
to S-Pbca structural transition drives the Mott transition 
associated with an orbital ordering~\cite{nakatsuji} across 356 K. However, Gorelov et al., do not get any
orbital selective Mott transition for the L-Pbca phase within their analysis down to 300 K. 
After wannierization of the GGA band structure for L-Pbca, we employ local correlations within DMFT+CT-QMC framework 
and try to see if there is Mott transition somewhere at lower temperature.
Subsequently the questions of interest are; what is the nature of the Mott insulator? Re-paraphrased somewhat, whether all the active
orbitals become Mott insulating or there is orbital selectivity? How do the DMFT results compare with
available ARPES results? And finally, is the desired Mott Insulating state recovered at 356K
if the S-Pbca structure is subjected to local correlations within GGA+U?  We choose $U=3.1 eV$ and $J_{H}=0.7 eV$ for the present analysis~\cite{pavarini}. As the temperature is lowered from  
1000K to 360K, we observe that unlike the rest of the materials that we have studied, there is no monotonic
tendency for Im$\Sigma(i\omega_{n})$ and Im$G(i\omega_{n})$ (Fig~\ref{fig9}(b)) towards coherence. Rather, orbital specific
loss of coherence can be clearly observed by lowering the temperature~\cite{arpes2,arpes1}. Finally it is found that the $d_{xy}$
orbital becomes Mott gapped with consistent singular low energy features in the Im$\Sigma(i\omega_{n})$
below 250K. The opening of the charge gap is clearly orbital selective~\cite{arpes1} in nature, 
as other orbitals remain metallic as far as their single-particle dynamic responses are concerned, down to the lowest temperature. One important point to note is that for $Ca_{1.5}Sr_{0.5}RuO_{4}$, the orbital with primarily $d_{xy}$ character was the one that became pseudogapped and for $x=2.0$ the orbital with major contribution from $d_{xy}$ becomes Mott gapped below 250K (Fig~\ref{fig9}(c)).
The critical Mott temperature, as is apparent, is much lower than the experimentally realised 356K scale.
ARPES studies do suggest~\cite{arpes1,arpes2} that the transition is orbital selective in nature, although the absence of orbital-selective Mott nature is also supported by one ARPES study~\cite{arpes3} and one theoretical study~\cite{pavarini}. Not much can, however, be concluded from the intercept of Im$\Sigma(i\omega_{n})$ at $i\omega_{n}=0$ and the orbital specific mass enhancement factors (Fig~\ref{fig9}(d)), except for the fact that the intercept becomes extremely large in proximity of Mott transition. We find that Im$\Sigma(i\omega_{n})$ becomes singular below a critical Mott temperature, and the quasi-particle description becomes somewhat untenable. 

In the two-particle sector, $\chi_{s,loc}(\tau)$ (Fig~\ref{fig9}(e)) suggests that the local moment remains
unquenched at all energy scales in the temperature range between 800K to 400K. However, 
although spins remain singular down to the Mott critical temperature ($T_{c,Mott}$), 
the behavior just above 356 K is deviant from Curie-Weiss. The high temperature $\chi_{s,loc}(T)$ 
is strictly Curie-Weiss within our analysis though, and a $\chi^{-1}_{s,loc}(T)$ vs $T$ linear fit remarkably predicts $\theta_{c}$ to be negative (-23 K) (Fig~\ref{fig9}(f)), suggesting that the material is in proximity of a low temperature antiferromagnetic Mott instability. As is well know from the literature, the system does have an antiferromagnetic insulating ground state below 113 K. This, in fact, is a success of our local analysis which correctly predicts the 
\onecolumngrid

\twocolumngrid

\noindent local instabilities in the excited spectrum that can lead to ground states realized in experiments.

We would like to point out that it is not possible to properly trace a self-consistent Mott metal-insulator transition across 356K within our local analysis which does not include the structural transition~\cite{pavarini} at 356K, role of phonons and the likelihood of an orbital ordering~\cite{orb1,orb2,orb3}.
We start with a high temperature L-Pbca structure and cool the system down to 250K. We observe that 
there is a possibility of an orbital selective Mott transition around this temperature. Next we performed a 
GGA calculation with the S-PBCA structure and included local Hubbard correlations within the GGA+U framework. We find that there is a Mott transition around 340K. This Mott transition, via the structural transition, has already been studied in literature~\cite{pavarini}. It is interesting to compare the relative scales of local correlations $U$ and $J_{H}$ with the band-widths of the dispersive electronic orbitals around the Fermi level across the series. $Sr_{2}RuO_{4}$ has a larger bandwidth ($W$), roughly 2.8 eV, and the $U/W$ ratio is about 0.89, while $J_{H}/W$ is 0.18. The Mott criteria suggests that $U/W$ has to be of order one or more to facilitate a Mott transition. A comparatively large $J_{H}/W$ in addition takes the system away from the proximity to a Mott transition, as the system is less than half filled. For $Ca_{2}RuO_{4}$ the effective bandwidth of the bands crossing the Fermi level is only $\sim 1 eV$. The large distortion of the octahedra narrows the $d\, t_{2g}$ bands significantly and leads to increment in the $U/W$, driving the material to the proximity of a Mott transition via a structural transition. This effectively pushes the system away from the Hund's limit and puts it in the Mott limit. The L-Pbca to S-Pbca transition is again associated with a flattening of the octahedra with nearly $10\%$ decrease in $c/a$ ratio at 356 K. 
An important point to note here is the relative values of the in-plane and out-of-plane effective hopping 
for $Ca_{2}RuO_{4}$ (Fig~\ref{fig9}(a)). Figure~\ref{fig9}(a) clearly shows that, even inside the L-Pbca phase, 
just above the Mott transition, $c/a$ is $\sim 2.06$ and the out of plane hopping is significantly suppressed in comparison to the in-plane hopping scale, $t_{c}/t_{ab}$ ratio being 0.3 at $x=0$. The nearly $2D$ nature~\cite{nakatsuji} of the $Ca_{2}RuO_{4}$ bulk single crystal is in marked contrast to its $x=0.5$ counterpart, as far as the effective dimensionality (measured by $t_{c}/t_{ab}$ ratio $\sim 1$ and $c/a$ ratio 3.3 for $x=0.5$) is concerned.


Put together all these and the previous theoretical 
\pagebreak
\onecolumngrid

\begin{figure}
\subfigure[]{\label{f:C11}\epsfig{file=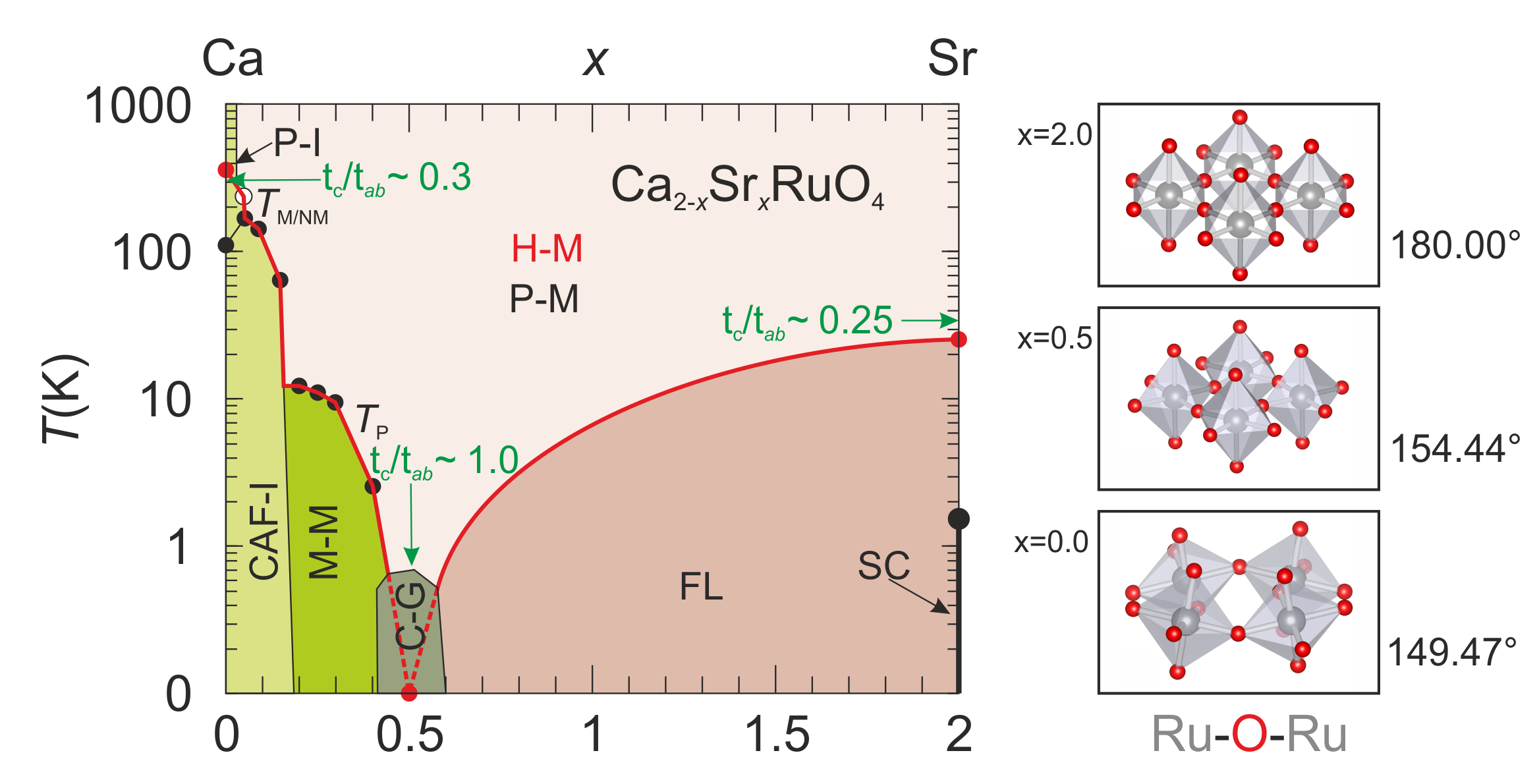,trim=0in 0in 0in 0.0in,
clip=true,width=0.88\linewidth}}\hspace{-0.0\linewidth}
\caption{Modified magnetic and electronic phase diagram for $Ca_{2-x}Sr_{x}RuO_{4}$ showing aspects of critical
dimensional crossover across $x=0.5$. The ratio of $t_{c}/t_{ab}$ across the phase diagram explicitly shows the aspects of
structural criticality at $x=0.5$. A Hund's metallic phase spans the quantum critical regime at finite $T$ and finite $x$ emanating from
$T=0, x =0.5$ end point. The singular enhanced spin fluctuation associates the critical end point. The critical fan suppresses a
low temperature Fermi liquid phase to the right $x>0.5$ and a magnetically order phase to its left $x<0.5$. The octahedral
distortions across the phase diagram and $Ru-O-Ru$ bond angles are shown in parallel panel.}
\label{fig10}
\end{figure}

\twocolumngrid

\noindent and experimental findings, we see that the $T = 0, x = 0.5$ is a quantum critical end point of a second-order structural transition separating two quasi $2D$-systems on both sides, to its right ($x > 0.5$) as well as left ($x < 0.5$) (Fig~\ref{fig10}). As far as crystal structures are concerned, the $x=2.0$ structure has higher crystal symmetries than the one at $x=0.0$. The dynamic in-plane susceptibility, its anisotropy and variation across the Brillouin zone (different momentum vectors) at different energy scales~\cite{terakura1,critical} also support the fact that an effective dimensional crossover attends the structural transition. Interestingly, the critical end point is also 
associated with a strong local ferromagnetic fluctuation extending to finite temperatures and $x$ away from 0.5.
The order of magnitude increment in local static susceptibility at this particular point of the phase diagram and the falling of~\cite{nakatsuji} on both left and right side of $x=0.5$, substantiate our claims. Our findings, supported amply by experimental results, therefore, raises an important question:  {\it is the critical point purely structural in nature?}.

At this moment we would again like to rely on the detailed calculations and the results discussed in our present work, and infer that, it is the structural change of the crystal via replacement of larger cations
with smaller cations that leads to this rich structural and magnetic phase diagram. Here, the structural changes are, therefore, the driving force behind the associated magnetic transitions and dimensional reductions. But these analyses of the microscopics of the fundamental crystal structures, 
rotations and hybridizations of the active bands at the Fermi level, role of dynamic correlations
on those bands, their single and two-particle (vertex-corrected) dynamic responses, allow us a fresh look into this important series with $K_{2}NiF_{4}$ structure, the structural building block of another interesting unconventional superconducting series. The structural changes,
\pagebreak
\onecolumngrid

\twocolumngrid 
\noindent the associated criticality, spin fluctuations and effective dimensional crossovers are used as fresh looking glasses into the iso-electronic series. However, it finally leads us to one interesting and relevant question: does the quantum critical end point of second-order structural transition lead to a quantum critical fan, much like what is realized in high $T_{c}$ unconventional superconductivity~\cite{subir,chu,ni} or heavy fermion compounds~\cite{mathur,si}?
Based on our analysis, we believe we have established a modified magnetic and electronic phase diagram for the series (Fig~\ref{fig10}). The paramagnetic bad metallic phase that emanates from this end point towards higher temperatures for all $x$ at and away from 0.5, extends all the way to the right till $x = 2.0$ and to the left till $x = 0.0$. At $x = 2.0$, above 25 K, the system is bad-metallic and at $x = 0.0$ the system is again bad- metallic above 356 K. This leads us to infer that the critical fan would be the one connecting the end point ($x = 0.5, T = 0$) to $x = 0.0, T = 25 K$ (to the right) and $x = 2.0, T = 356 K$ to the left. The iso-electronic material is a Hund's metal inside this critical fan, and outside, it is a good metal (Fermi-liquid) below 25 K at $x = 2.0$ and a magnetic material (either ferro or anti-ferro) for $x < 0.5$. It is possible only at $T=0$, that the $x > 0.5$ good-metal of 2D nature without any magnetic ground state can be tuned through the $3D$ critical end point at $x=0.5$ and made a magnetically ordered material (metal and insulator respectively at $0.2 < x < 0.5$ and $0 < x <0.2$) in the region $x < 0.5$. While our studies suggest that at finite temperatures $< 25 K$ a good metal for $x > 0.5$ can be tuned inside the critical fan and made a Hund's metal that finally crosses the other side of the fan at $x< 0.5$ and becomes 
an antiferromagnetic insulator. From our analysis, a strictly local one, the aspects of local fluctuations, criticality, Fermi liquidity and Hund's metallicity can be well described. However, what it does not describe are non-local low energy fluctuations that may suppress the access to the $T=0,\, x=0.5$ critical end point. Very recent experimental studies suggest that the $T=0,\, x=0.5$ point is cluster glass~\cite{carlo}, which is beyond the scope of our local analysis. That will lead to the modification of the phase diagram we have come up with, where the critical fan will end at the boundary of the cluster glass phase, and should not extend down to $T=0$. However, inclusion of the cluster glass phase does not forbid the microscopics of the finite temperature aspects of criticality and the structure of the critical fan sustaining a Hund's metallic phase. 

SA would like to acknowledge useful discussions on the first-principles calculations with Monodeep Chakraborty. 
SA acknowledges discussions with M S Laad and Debraj Choudhury and thanks Arjun Mukerji for help in preparing a diagram. SA acknowledges UGC (India) and DD acknowledges DST (India) for research fellowships.

\end{document}